\documentclass[10pt,conference]{./IEEEtran}
\usepackage{graphicx}
\usepackage{subfigure}
\usepackage{epstopdf}
\begin{document}

\title{Compressed Shattering}

%\author{
%\authorblockN{A. Chockalingam}
%\authorblockA{Dept.\ of Electrical Communication Engg.\\
%Indian Institute of Science \\
%Bangalore, 560012, India\\
%Email: spcom2016@gmail.com} \and
%\authorblockN{Rahul Vaze}
%\authorblockA{School of Technology and Computer Science\\
%Tata Institute of Fundamental Research\\
%Mumbai, 400005, India\\
%Email: spcom2016@gmail.com}
%\and
%\authorblockN{Animesh Kumar}
%\authorblockA{Dept.\ of Electrical Engg.\\
%Indian Institute of Technolgy \\
%Bombay, 400076, Mumbai India\\
%Email: spcom2016@gmail.com}
%%\and
%%\authorblockN{Robert Calderbank}
%%\authorblockA{Department of Electrical Engineering\\
%%Princeton University \\
%%Princeton, NJ 08544, USA\\
%%Email: calderbk@math.princeton.edu} \and
%%\authorblockN{Habong Chung}
%%\authorblockA{Department of Electronic\\
%%\& Electrical Engineering\\
%%Hongik University\\
%%Seoul, Korea\\
%%Email: habchung@hongik.ac.kr}
%}
% avoiding spaces at the end of the author lines is not a problem with
% conference papers because we don't use \thanks or \IEEEmembership

%%%%\author{Nithin Nagaraj$^1$, Harikumar K., Anand K. N., Sandra K., Shreeja Sugathan} %\IEEEauthorrefmark{1}
\author{
\IEEEauthorblockN{Harikumar Kannampillil, Anand Krishnadas Nambisan, Sandra Kizhakkekundil, Shreeja Sugathan}
     \IEEEauthorblockA{Amrita School of Engineering, Amrita Vishwa Vidyapeetham, Amritapuri, Kollam, Kerala, INDIA.}
     \IEEEauthorblockN{Nithin Nagaraj}
     \IEEEauthorblockA{School of Humanities, National Institute of Advanced Studies, \\ Indian Institute of Science Campus, Bengaluru, INDIA. (Email: nithin@nias.iisc.ernet.in)}}
    %\\\{2, 3\}@def.com}

%\author{
%\authorblockn{Nithin Nagaraj, Harikumar K., Anand K. N., Sandra K., Shreeja Sugathan }
%\authorblocka{Department of Electronics and Communication Engg.\\
%Amrita School of Engg., Amrita Vishwa Vidyapeetham \\
%Amritapuri Campus, Kollam, Kerala 690525 INDIA. \\
%nithin@am.amrita.edu
%}
%}

\maketitle

\begin{abstract}
The central idea of compressed sensing is to exploit the fact that most signals of interest are sparse in some domain and use this to reduce the number of measurements to encode. However, if the sparsity of the input signal is not precisely known, but known to lie within a specified range, compressed sensing as such cannot exploit this fact and would need to use the same number of measurements even for a very sparse signal. In this paper, we propose a novel method called Compressed Shattering to adapt compressed sensing to the specified sparsity range, without changing the sensing matrix by creating shattered signals which have fixed sparsity. This is accomplished by first suitably permuting the input spectrum and then using a filter bank to create fixed sparsity shattered signals. By ensuring that all the shattered signals are utmost 1-sparse, we make use of a simple but efficient deterministic sensing matrix to yield very low number of measurements.  For a discrete-time signal of length 1000, with a sparsity range of $5 - 25$, traditional compressed sensing requires $175$ measurements, whereas Compressed Shattering would only need $20 - 100$ measurements.
\end{abstract}

\begin{figure*}[!ht]
	\centering
	\includegraphics[width=5.3in]{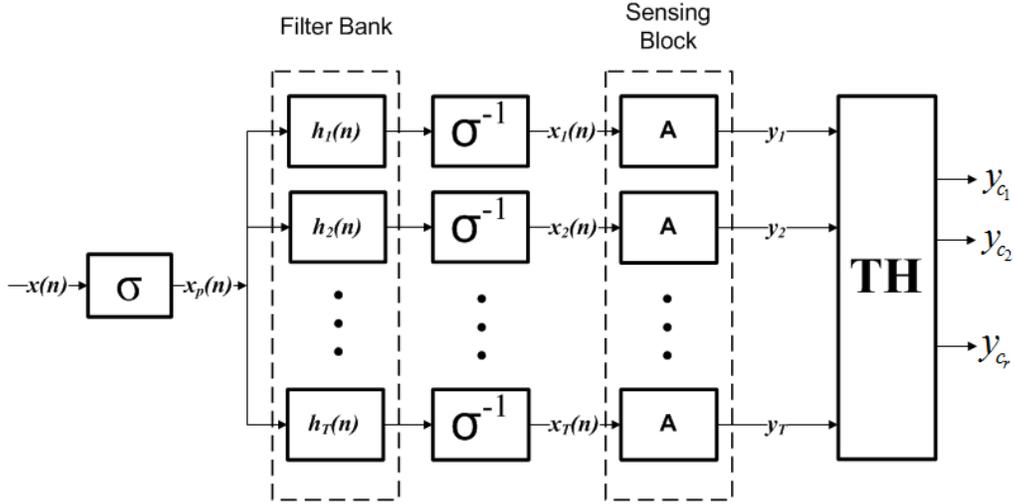} %5.9 in
	\caption{Block diagram which illustrates the compressed shattering process. $\sigma$ and $\sigma^{-1}$ represent the permutation operation with the respective parameters, {$h_{1}(n), h_{2}(n), \ldots, h_{T}(n)$} are the impulse responses of $T$ filters of the filter bank, {$x_{1}(n), x_{2}(n), \ldots, x_{T}(n)$} are the outputs of the $T$ filters after inverse permutation, {$y_{1}, y_{2}, \ldots, y_{T}$} are the measurements corresponding to the $T$ filters, \textbf{TH} represents the threshold operation block, {$y_{c_{1}}, y_{c_{2}}, \ldots, y_{c_{r}}$} represent the $r$ significant outputs.}
	\label{forward_compresss_filtering}
\end{figure*}

\begin{figure*}[!ht]
	\centering
	\includegraphics[width=5.3in]{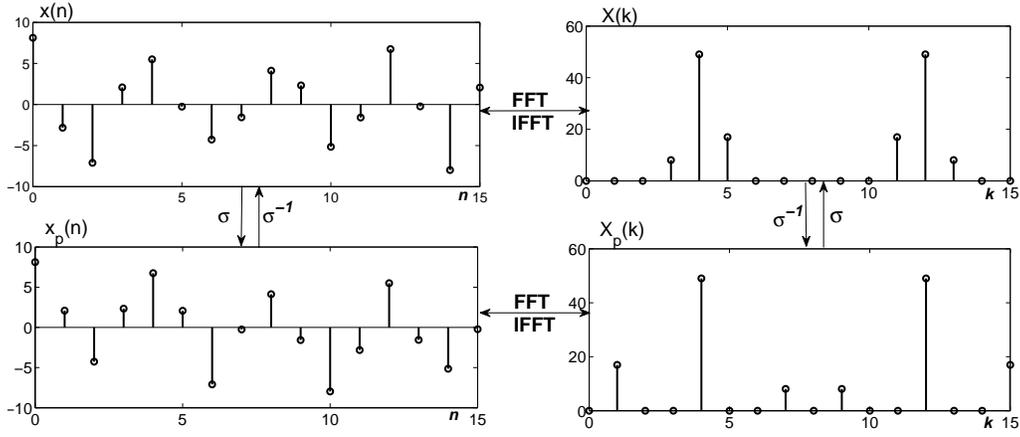} %5.9
	\caption{Permutation block: $N=16$, $m=3$, $\sigma=3$, $\sigma^{-1}=11$, $n$ represents time domain, $k$ represents frequency domain, $x(n)$ is the input signal, $x_{p}(n)$ is the permuted signal, $X(k)$ is the input signal spectrum, $X_{p}(k)$ is the permuted signal spectrum.}
	\label{permutation_unclustering}
\end{figure*}

\begin{figure*}[!ht]
	\centering
	\includegraphics[width=6.4in]{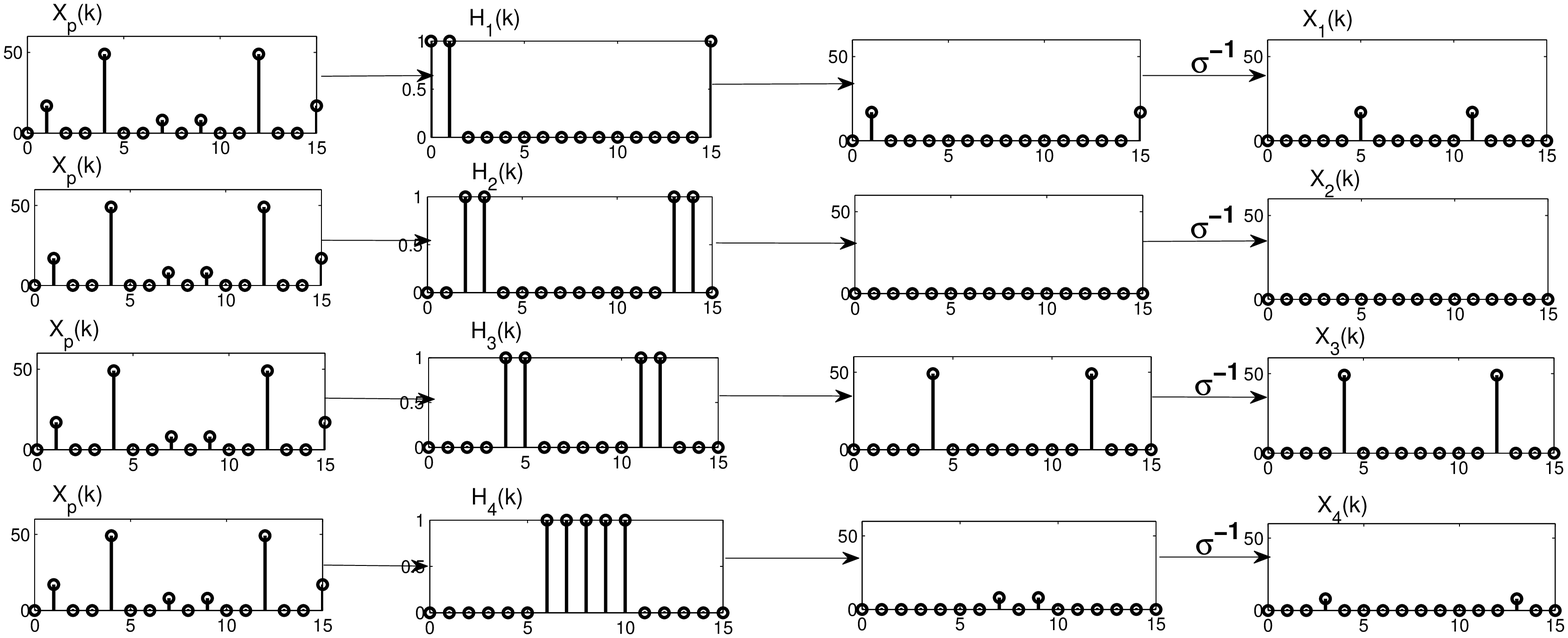}
	\caption{$N=16$, $m=3$, $T=4$, $\sigma=3$, $\sigma^{-1}=11$, $X_{p}(k)$ is the same permuted spectrum from Fig. \ref{permutation_unclustering}. $H_{g}(k)$ is the filter response of the $g^{th}$ filter and $X_{g}(k)$ is the spectrum after de-permutation operation from the $g^{th}$ filter where $g = 1, 2, \ldots, T$. This figure shows what happens in the spectral domain, when the permuted signal $x_{p}(n)$ is passed through through the corresponding filters and gets de-permuted. Note that the operation with $\sigma^{-1}$ is done in time domain.}
	\label{filter_performance_mega}
\end{figure*}

\section{Introduction}
Compressed sensing~\cite{Candes} is a fundamental idea in mathematics, which utilizes the {\it a priori} property of signal $x(n)$ of length $N$ being $m$ sparse in some domain, where $m<<N$, along with an appropriately constructed sensing matrix $A$, to establish a unique solution for an otherwise undetermined system of linear equations:
\begin{equation}
A_{M\times N}\cdot x_{N\times 1}=Y_{M\times 1}.
\end{equation}
The actual solution, is the vector from the solution set, which has the minimum \(l_{0}\) norm. Since this is a NP-hard problem, so we choose the solution which minimizes the \(l_{1}\) norm. It is observed that minimizing \(l_{1}\) norm will give an accurate solution~\cite{Tao} provided the sensing matrix $A$ satisfies the Restricted Isometry Property ($RIP$) property~\cite{Candes}.

However, if the sparsity of the input signal is not precisely known, but known to lie within a specified range, traditional compressed sensing as such cannot exploit this fact and would need to use the same number of measurements for all sparsity values in this range. In this case, the compressed sensing algorithm has to work taking into account the worst case, which corresponds to the signal being least sparse. For example if the input signal is a discrete-time digital signal of length $N$ and can have sparsity anywhere between $m_1=5$ to $m_2 =25$ in the frequency domain, for compressed sensing to work, one has to design the sensing matrix keeping in mind the sparsity value $m_2=25$. For this case, there are $25$ frequencies in the signal which will correspond to $48 - 50$ complex coefficients (depending upon the locations of those 25 frequency coefficients), it was experimentally observed to take about $175$ ($=7 \cdot 25$) measurements for an accurate reconstruction by minimizing the $l_{1}$ norm. Thus if the input signal had sparsity $m=5$, conventional compressed sensing would take $175$ measurements (since it has been designed for sparsity $m_2=25$) whereas only 40 measurements would have sufficed. Thus, we have unnecessarily used 135 more measurements than needed in this case.

In this paper, we propose a novel method called Compressed Shattering to address this particular issue. The central idea of compressed shattering is to adapt compressed sensing to the specified sparsity range by creating shattered signals~\cite{Gilbert} which have fixed sparsity using a filter-bank. Our primary aim is to reduce the number of measurements.

\section{Compressed Shattering}
The problem is stated as follows. The input signal is a discrete-time digital signal of length $N$ which needs to be sensed. It is sparse within a range $\left[m_1,m_2\right]$ in the frequency domain, where $0\leq m_1\leq m_2\leq \left\lfloor N/2\right\rfloor$. Here,  $m$-sparse means there are only $m$ non-zero coefficients in the $DFT$ of the input signal, without considering the symmetric complex conjugate parts.
%As an example, if $N=1000$ and $m=6$, only ?? are non-zero ??.

\subsection{Block Diagram: Overview}
The proposed algorithm is described in Fig.~\ref{forward_compresss_filtering}. First the input signal is permuted. This results in a permutation of the spectrum ($DFT$) in order to remove any clusters and to spread it out. The permuted spectrum is then passed through a filter-bank, which is a set of \(T\) band-pass filters, where \(1\leq T\leq \left\lfloor N/2\right\rfloor\). An inverse permutation operation is done on all the filter outputs to put the spectrum back in its original position. The compressed sensing algorithm is applied on the output of the filters; using the same sensing matrix on each of them.  Depending upon the sparsity level of the original signal, the filter outputs might be zero or a very sparse signal, and the level of sparsity in the output of each of the filters can be controlled by adjusting the characteristics and number of filters. In the succeeding subsections, we describe each block in detail.
%
%

%Permutation, being a linear operation, can be seen as a multiplication between the input vector and the corresponding permutation matrix. An \(N\) length %discrete signal can be permuted in \(N!\) ways.  The \(N\times N\) permutation matrix is invertible and has rank \(N\). But here, from the set of all possible %permutation matrices, we are interested in a subset, which results in the operation specified in Eqn. (1):

\subsection{Permutation Block}
The permutation block performs a mapping operation in which the indices of input signal are rearranged. It is given by:
\begin{equation}
x_{p}\left(n \right)= x((\sigma \cdot n)~mod\;N),
\end{equation}

where \(x(n)\) is the input discrete-time signal of length \(N\) and $n = 0, 1, 2, \ldots, N-1$. It is to be noted that all the operations performed on the indices are modulo \(N\) operations. The parameter \(\sigma\) should be relatively prime to \(N\) to ensure that the resultant permutation matrix is invertible. Considering \(N\) to be power of \(2\), any odd number belong to set \({1, 3, \ldots, N-1}\) would suffice. Permutation done using  $\sigma$ will ensure that the spectrum of the signal also gets permuted but with \(\sigma^{-1}\) as the permutation parameter~\cite{Gilbert}, in accordance with the equation:
\begin{equation}
X_{p}\left(k \right)= X((\sigma^{-1} \cdot k)~mod\;N),
\end{equation}
where \(X(k)\) is the $N$-$DFT$ of $x(n)$ and $k = 0, 1, 2, \ldots, N-1$ and $\sigma^{-1}$ is defined as $(\sigma\cdot\sigma^{-1})~mod~N = 1$.  Fig. \ref{permutation_unclustering} depicts the permutation on an example, it can also be seen that this permutation helps to de-cluster the signal spectrum\footnote{The permutation operation can be seen as a linear congruential generator which randomizes the indices. Instead, a more powerful pseudo-random number generator (PRNG) could be used.}.

\subsection{Filtering and Inverse Permutation}
The permuted signal is then passed through a filter-bank of $T$ non-overlapping ideal filters. It should be noted that the filter design and the algorithm that follows in this paper is done by considering $N$ to be even and the number of filters $T$ divides $\frac{N}{2}$ to give an integer. The frequency response of the filter banks are:
\begin{equation}
\setlength{\nulldelimiterspace}{0pt}
H_{b}(k)=\left\{\begin{IEEEeqnarraybox}[\relax][c]{ls}
1,\;\;\;\;\;\;& $\frac{(b-1)\times N}{2\times T}\leq k< \frac{N\times b}{2 \times T}$\\
1,\;\;\;\;\;\;& $ a_1< k\leq a_2$\\
0,\;\;\;\;\;\;&elsewere%
\end{IEEEeqnarraybox}\right.,
\end{equation}
where,
\begin{equation}
a_1 = N-\frac{b\times N}{2\times T},
\end{equation}
\begin{equation}
a_2 = N-\frac{(b-1)\times N}{2\times T},
\end{equation}
for $b = {2,3, \ldots, (T-1)}$. When $b=1$, we have: $H_{1}(k)$, where $a_2=N-1$. When $b = T$, we have:

\begin{equation}
H_{T}(k)\;,\;H_{T}(\frac{N}{2}) = 1.
\end{equation}

Out of the $T$ filters, only $r$ of them will have significant outputs,  where $1\leq r\leq m_2$. This filter bank will do a circular convolution as opposed to the normal linear convolution. It is done to preserve the length of the signal and it will be a perfect element wise multiplication in the Fourier domain without having to pad any zeros. Preservation of length is necessary for the inverse permutation block that comes next. The signal is then passed through it for reversing the permutation operation, thereby putting the spectrum back in its original position. Fig. \ref{filter_performance_mega} gives an example of filtering and inverse permutation operations.

The output signals obtained at the end of filtering and inverse permutation are known as {\it shattered} signals~\cite{Gilbert,Gilbert2}. As can be seen, the shattered signals are relatively more sparse than the input signal and the sparsity can be controlled by suitably changing $\sigma$ and $\{ H_b(\cdot) \}$. It is also possible to obtain shattered signals which are at most 1-sparse as the outputs.

\subsection{Sensing Block}
The shattered signals are now ready for compressed sensing. Each of the output signals are sensed by the same sensing matrix $A$, designed for specific level of sparsity, which will preferably be much less than the original minimum sparsity  $m_2$ of the signal specified in the range. Note that more the number of frequencies present in the signal, the less sparse is its spectrum. Although it is possible to obtain different level of sparsity for the shattered signal, in this paper, we have ensured that the shattered signals are all 0 or 1-sparse. In other words, the filter outputs have at most a single frequency. Hence the number of filters $T$ should be at least $m_2$.  This is sensed by a $2 \times N$ sensing matrix $A$ specifically designed to sense such 1-sparse data, taking into account the symmetry of the $DFT$. By this, we ensure that each of the non-zero shattered signals can be sensed in just $2$ measurements. In total that will amount to at most $2\cdot T$ measurements.

\subsection{Deterministic Sensing Matrix $A$}
As opposed to use of a sensing matrix with random values, we propose a simple but efficient deterministic sensing matrix $A$. We make use of the information that shattered signals are either 0 or 1-sparse. The number of unknowns are just two for each output (position and value of complex $DFT$ coefficient). We also make use of the fact that the $DFT$, for real signals, is conjugate symmetric.

\begin{equation}
A_{2\times N}\cdot x_{g}=y_{g},
\end{equation}

\begin{equation}
A_{2\times N} = \Phi_{2\times N}\cdot \Psi_{N\times N},
\end{equation}

\begin{equation}
\Psi_{N\times N}(k,n) = e^{\frac{-i2\pi nk}{N}} ,
\end{equation}

\begin{equation}
\Phi_{2\times N} = \left(\begin{IEEEeqnarraybox*}[][c]{,c/c/c/c/c/c/c/c,}
\cos(\theta_{0})&\cos(\theta_{1})&$...$&\cos(\theta_{\frac{N}{2}})&0&0&...&0\\
\sin(\theta_{0})&\sin(\theta_{1})&$...$&\sin(\theta_{\frac{N}{2}})&0&0&...&0
\end{IEEEeqnarraybox*}\right),
\end{equation}

where $\theta_{s} = \frac{\pi \times s}{N}$ ($s={0, 1, \ldots, \frac{N}{2}}$), $x_{g}$ is the signal of length $N$ at output of the $g^{th}$ filter ($g=1, 2, \ldots, T$) which has at most a single frequency (0 or 1-sparse). $\Psi$ is the $N-DFT$ matrix, $A$ is the sensing matrix of which any two columns of the first $\frac{N}{2}$ columns of A are linearly independent, $y_{g}$ is $(2\times 1)$ measurement vector which is complex valued. This sensing matrix will ensure that the sensing happens only for the first half of the spectrum. Further, at most only $r$ of the filters will have significant output, namely ${c_{1},c_{2}, \ldots, c_{r}}$. So, only the measurements, $y_{c_{j}}$ ($j = 1, 2, \ldots, r$), corresponding to those $r$ filters needs to be stored. For this reason all the measurements $\{ y_g\}$,  is passed through a threshold block \textbf{TH} (refer to Fig.~\ref{forward_compresss_filtering}), where insignificant measurements are discarded by choosing an appropriate threshold for the $l_2$ norm of the shattered signals.

\section{Reconstruction Block}

\begin{figure}[!ht]
	\centering
	\includegraphics[width=3.3in]{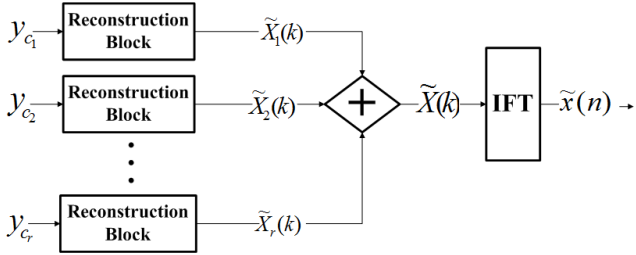} %3.4in
	\caption{Block diagram which illustrates the process of reconstruction.}
	\label{Reconstruction_compress_filtering}
\end{figure}

\begin{table*}[!ht]
	\centering
	\caption[Comparative Results]{Simulation Results for sparsity values $m=5$ and 25.} %
	\label{tab:tableresults} %
	\vspace{0.01in}
	%\begin{tabular}{|@{}c@{}|@{}c@{}|@{}c@{}|@{}c@{}||@{}c@{}|@{}c@{}||@{}c@{}|@{}c@{}||@{}c@{}|}
	%\begin{tabular}{|c|c|c|c||c|c||c|c||c|}
	\begin{tabular}{|@{}c@{}|c|@{}c@{}|@{}c||c@{}|@{}c||c@{}|@{}c||c@{}|}
		\hline
		&  &   & \multicolumn{2}{|c|}{No. of Measurements}   & \multicolumn{2}{|c|}{No. of Additions} & \multicolumn{2}{|c|}{No. of Multiplications}\\
		\cline{4-5} \cline{6-7} \cline{8-9}
		$N$    & $m$    & $T$  & Compressed    & Compressed       & Compressed  & Compressed & Compressed   &  Compressed \\
		
		&        &      & Sensing  & Shattering      &     Sensing &      Shattering &      Sensing &     Shattering\\
		\hline
		1000 & 5 &  100       &  175  & 20   & 174825 & 399600 & $1.75 \times 10^5$ & $4\times 10^5$\\
		1000 & 25 & 100      &  175 & 100 & 174825 & 399600 & $1.75 \times 10^5$ & $4\times 10^5$\\
		\hline
	\end{tabular}
\end{table*}

In compressed shattering, since we are using a deterministic sensing matrix $A$ as described above, we can make use of the inherent structure in the matrix to design a very fast reconstruction algorithm. We directly calculate the position and the value of the frequency coefficient of the signal by the following set of equations: (because $\theta_{s} \leq \frac{\pi}{2}$)
\begin{equation}
\Theta_{j} = \cos^{-1}(\frac{\left| y_{c_{j}}(0)\right|}{||y_{c_{j}}||_2}),
\end{equation}

\begin{equation}
\alpha_{j} = \frac{\Theta_{j}}{\Delta \theta}\;\;,\;\;\beta_{j} = \frac{y_{c_{j}}(0)}{\cos(\Theta_{j})},
\end{equation}

%\begin{equation}

%\end{equation}

\begin{equation}
\Delta \theta = \frac{\pi}{N},
\end{equation}

where $\beta_{j}$ represents the complex coefficient and $\alpha_{j}$ represents the position of the coefficient. From the above equation we can reconstruct the spectrum of the signal in the following way. (When $\alpha_{j}\neq 0$)

\begin{equation}
\setlength{\nulldelimiterspace}{0pt}
\widetilde{X}_{j}(k)=\left\{\begin{IEEEeqnarraybox}[\relax][c]{ls}
\beta_{j},\;\;& $k = \alpha_{j}$\\
\beta_{j}',\;\;& $k = N-\alpha_{j}$\\
0,\;\;\;\;\;\;\;\;\;\;\;\;\;\;\;\;\;&elsewere%
\end{IEEEeqnarraybox}\right.\;\;\;,
\end{equation}

when $\alpha_{j} = 0$,

\begin{equation}
\setlength{\nulldelimiterspace}{0pt}
\widetilde{X}_{j}(k)=\left\{\begin{IEEEeqnarraybox}[\relax][c]{ls}
\beta_{j},\;\;& $k = 0$\\
0,\;\;\;\;\;\;\;\;\;\;\;\;\;\;\;\;\;&elsewere%
\end{IEEEeqnarraybox}\right.,
\end{equation}

where $\beta_{j}'$ is the complex conjugate of $\beta_{j}$ and $\widetilde{X}_{j}$ is the reconstructed spectrum of the output of the filter $c_{j}$. Summing up all respective reconstructed spectrums of the $r$ significant filters will give the reconstructed version of the original signal spectrum represented by $\widetilde{X}$ (refer to Fig.~\ref{Reconstruction_compress_filtering}).
\begin{equation}
\widetilde{X} = \sum_{j=1}^{r}\widetilde{X}_{j},
\end{equation}

\begin{equation}
\widetilde{x}(n) = \sum^{N-1}_{k = 0}\widetilde{X}(k)\cdot e^{\frac{i2\pi nk}{N}}.
\end{equation}

By taking the Inverse $DFT$ of $\widetilde{X}(k)$ we get $\widetilde{x}(n)$ which represents the reconstructed version of the original time domain signal.

\section{Matrix Formulation}
To summarize, compressed shattering has four steps in the following order.  The input signal is 1) permuted, 2) passed through a filter-bank, 3) de-permuted, and 4) finally sensed by a sensing matrix $A$. There will be $T$ such paths corresponding to $T$ filters, however only $r$ will be significant (refer to Fig.~\ref{forward_compresss_filtering}). Since every block is a linear transformation (up to the thresholding block), we can reduce the entire compressed shattering procedure to one single matrix (for each of the $T$ paths). This is given by:
\begin{equation}
A\cdot x_{g} = y_{g}
\end{equation}
here $x_{g}$ can be replaced with the following:
\begin{equation}
A_{2\times N}\cdot P^{-1}_{N\times N}\cdot Hmat_{g}\cdot P_{N\times N}\cdot x_{N \times 1} = y_{g},
\end{equation}

where $P$ is the permutation matrix, $P^{-1}$ is the inverse permutation matrix and $Hmat_{g}$ is an $(N \times N)$ circular convolution matrix corresponding to the $g^{th}$ filter. These matrices can be multiplied to form a single matrix $\gamma_{g}$, of size $(2\times N)$, with complex entries, that takes the input $x$ and transforms it into the measurements $y_{g}$ corresponding to the $g^{th}$ filter:

\begin{equation}
\gamma_{g}\cdot x= y_{g},
\end{equation}
\begin{equation}
\setlength{\nulldelimiterspace}{0pt}
\left[ \begin{array}{c} \gamma_{1}\\\gamma_{2}\\.\\.\\.\\\gamma_{T}\end{array} \right]_{2T\times N}\cdot\;\;\;\;\; x_{N \times 1}\;\; = \left[ \begin{array}{c} y_{1}\\y_{2}\\.\\.\\.\\y_{T}\end{array} \right]_{2T\times 1}.
\end{equation}
%
%\begin{equation}
%\Gamma_{2T\times N}\cdot x = Y_{2T\times 1}
%\end{equation}

\section{Simulation Results and Discussion}
In this section, we perform numerical simulations to test our proposed algorithm and compare it with conventional compressed sensing. The parameters for comparison will be number of measurements stored and number of computations. The input signal to the system is a discrete-time real signal of length $N=1000$ and will have sparsity anywhere in the range $m_1=5$ to $m_2=25$ frequencies. We report results for both the extreme cases of sparsity: $m_1$ and $m_2$.

\begin{figure}[!ht]
	\centering
	\includegraphics[width = 3.4in]{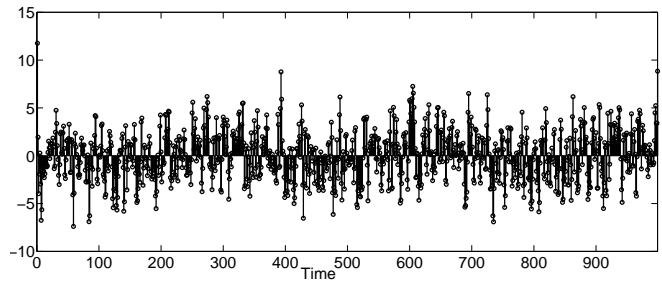}
	\caption{The input real valued signal $x(n)$ of length $N=1000$, which has a sparsity of $m_2 = 25$ frequencies. }
	\label{input_time_f25}
\end{figure}

\begin{figure}[!ht]
	\centering
	\includegraphics[width = 3.4in]{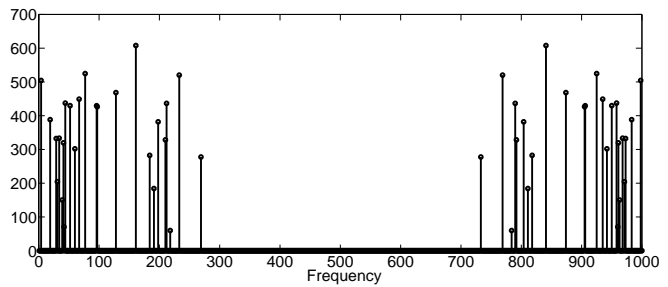}
	\caption{Absolute value of the spectrum $X(k)$ of the input shown in Fig. \ref{input_time_f25}.}
	\label{input_spectrum_f25}
\end{figure}

The input signal and its DFT spectrum corresponding to sparsity $m_2 = 25$ are shown in  Fig.~{\ref{input_time_f25}} and Fig.~{\ref{input_spectrum_f25}} respectively\footnote{We have omitted plotting the corresponding graphs for the signal with sparsity $m_1=5$ owing to space constraints.}. Table~\ref{tab:tableresults} shows the comparison between compressed sensing and compressed shattering in terms of number of measurements to be stored and number of additions and multiplications. Although $T=100$ filters are used in the compressed shattering algorithm ($\sigma$=11), very few shattered signals have significant energy indicating that most of them are 0-sparse. By choosing a threshold of 0.01 for the $||y_g||_2$, only very few shattered signals are retained as 1-sparse output signals. The measurements for compressed shattering are complex values whereas as compressed sensing yields real measurement values. However, in the table we have indicated number of real measurements which implies that we have multiplied the number of measurements for compressed shattering by 2. In all cases\footnote{We omit displaying the reconstructed outputs owing to space constraints.}, we obtained near-perfect reconstruction since the maximum absolute reconstruction error was $ < 10^{-11}$.

From the table, we can infer that there is a tradeoff between the number of measurements that have to be stored and the computational complexity involved in taking the initial measurement. Only half the number of real values have to be stored in the case of compressed shattering compared to the conventional compressed sensing method, but the computational complexity of the former is a little more than twice that of the latter in terms of both number of addition and multiplication. This is the price we pay for the reduction in number of measurements. It also should be noted that the algorithm, as of now, is heavily dependent on the $\sigma$ we choose.  So if we choose the wrong $\sigma$ the algorithm might fail because one of the filters might pick up more than one frequency.

\begin{figure}[!ht]
	\centering
	\includegraphics[width = 3.4in]{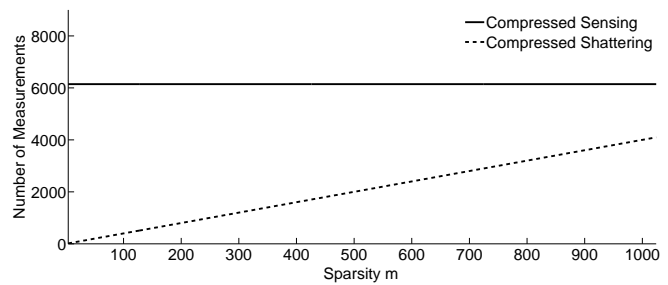}
	\caption{The graph shows a comparison of the number of measurements stored, for both compressed sensing and compressed shattering,  when subjected to the same input of length $2^{14}$ which has $m$ frequencies and $m$ is varied from $4 - 1024$. The number of filters used in the compressed shattering algorithm is $T=2048$. The number of measurements to be stored by compressed shattering is $(4 \cdot m)$, while for compressed sensing it is always fixed at 6144 ($=6 \cdot 1024$). Here, the gains provided by compressed shattering over compressed sensing method can be clearly observed, especially for small $m$.}
	\label{measurment_vs_m_N16384}
\end{figure}

A plot of number of measurements stored versus the sparsity $m$ is shown in Fig.~\ref{measurment_vs_m_N16384}, for $N = 2^{14}$.  The flexibility of compressed shattering to the sparsity range is evident when compared to traditional compressed sensing and thus results in huge gains, especially when sparsity $m$ is small.
%
%\begin{figure}[!ht]
%\centering
%\includegraphics[width = 3.4in]{reconstructed_spectrum_f25.eps}
%\caption{Reconstructed signal spectrum $\widetilde{X}$. This is almost exactly same as the input signal spectrum}
%\label{reconstructed_spectrum_f25}
%\end{figure}

%\begin{figure}[!ht]
%\centering
%\includegraphics[width = 3.4in]{error_time_f25.eps}
%\caption{Absolute error between the original signal and the reconstructed signal $\left|\widetilde{x}-x\right|$.}
%\label{error_time_f25}
%\end{figure}

%\begin{figure}[!ht]
%	\centering
%	\includegraphics[width = 3in]{measurment_vs_m_N1000.eps}
%	\caption{The graph shows a comparison of the number of measurements stored, for both compressed sensing and compressed shattering,  when subjected to the same input of length $1000$ which is has $m$ frequencies and $m$ is varied from $5 - 25$. The number of measurements here means the total number of real values to be stored.}
%	\label{measurment_vs_m_N1000}
%\end{figure}

%
%\begin{figure}[!ht]
%\centering
%\includegraphics[width = 3.4in]{error_time_f5.eps}
%\caption{Absolute error between the original signal and the reconstructed signal $\left|\widetilde{x}-x\right|$ where $x$ is the input with $5$ frequencies.}
%\label{error_time_f5}
%\end{figure}

%\section{Proceedings}
%All the accepted papers will be published in a conference
%proceeding. All the registrants at the conference will receive a
%copy of the proceedings.

\section{Conclusions and Future Research Work}
We have proposed Compressed Shattering - a novel way of extending compressed sensing when the sparsity of the input signal is within a specified range. The idea of using a linear congruential generator on the discrete-time indices helps to randomize the frequency components, and thus in de-clustering the spectrum. This is then exploited by creating 1-sparse signals by means of a filter-bank. Reconstruction  is very fast owing to a simple deterministic sensing matrix that we have proposed for 1-sparse signals. It is conceivable that a more sophisticated PRNG could be used to efficiently de-cluster the spectrum. Compressed Shattering outperforms traditional compressed sensing in terms of number of measurements that needs to be stored but at the cost of increased computational cost. Future research directions include studying compressed shattering in the presence of noise, finding optimal choices for $\sigma$, an enhanced PRNG, and a faster algorithm for generating shattered signals.

%\section*{Acknowledgment}
%We thank - if there is none specifically, we can comment this


\begin{thebibliography}{1}
	
	
	\bibitem{Candes}
	E.~J.~Candes and M.~B.~Wakin, ``An Introduction to Compressive Sampling,''
	\emph{IEEE Signal Processing Magazine}, vol.\ 114, pp.~21--30, 2008.
	
	\bibitem{Tao}
	E.~J.~Candes and T.~Tao, ``Near-Optimal Signal Recovery From Random Projections: Universal Encoding Strategies? ''
	\emph{IEEE Transactions on Information Theory}, vol.\ 52, pp.~5406--5425, 2006.
	
	
	\bibitem{Gilbert}
	J.~A.~T.~C.~Gilbert, M.~J.~ Strauss, ``A Tutorial on Fast Fourier Sampling,''
	\emph{IEEE Signal Processing Magazine}, vol.\ 25, pp.~57--66, 2008.
	
	\bibitem{Gilbert2}
	A.~C.~Gilbert, S.~Muthukrishnan, M.~Strauss, ``Improved Time Bounds for Near-Optimal Sparse Fourier
	Representations,''
	\emph{in Proc. SPIE Wavelets XI}, M.~Papdakis, A.~F.~Laine, and M.~A.~Unser, Eds., San Diego, CA, 2005, pp.~59141A.115.
	
	
	
	%
\end{thebibliography}
\end{document}